\documentclass[a4paper]{jpconf}
\usepackage{epsfig, amsmath, amssymb}
\usepackage{graphicx,psfrag} 

\newcommand{\be}{\begin{equation}}
\newcommand{\ee}{\end{equation}}
\newcommand{\bea}{\begin{eqnarray}}
\newcommand{\eea}{\end{eqnarray}}
\newcommand{\bA}{\begin{array}}
\newcommand{\eA}{\end{array}}
\newcommand{\bc}{\begin{center}}
\newcommand{\ec}{\end{center}}

\newcommand{\ra}{\rightarrow}
\newcommand{\del}{\partial}

\newcommand{\ie}{{\it i.e.}}
\newcommand{\eg}{{\it e.g.}}

\newcommand{\Nf}{${\cal N}{=}4$}

\newcommand{\lA}{\langle}
\newcommand{\rA}{\rangle}
\def\BC{{\mathbb C}}

\def\BZ{{\mathbb Z}}

\begin{document}





\title{Cosmological singularities, $AdS/CFT$ and de Sitter deformations}

\author{K~Narayan}
\address{Chennai Mathematical Institute, SIPCOT IT Park, Padur PO, 
Siruseri 603103, India.}


\begin{abstract}
We review aspects of certain time-dependent deformations of $AdS/CFT$
containing cosmological singularities and their gauge theory duals. 
Towards understanding these solutions better, we explore similar 
singular deformations of de Sitter space and argue that these 
solutions are constrained, possibly corresponding to specific
initial conditions. This is based on a talk given at the ICGC 2011 
conference, Dec 2011.

\end{abstract}




\section{Cosmological singularities and $AdS/CFT$}

General relativity breaks down at cosmological singularities, with
curvatures and tidal forces typically diverging: notions of spacetime
thus break down. There is a rich history of string theory explorations 
of such singularities \cite{stringSingRev}. We focus here on describing 
certain time-dependent deformations of AdS/CFT 
\cite{Das:2006dz,Das:2006pw,Awad:2007fj,Awad:2008jf}, where
the bulk gravity theory develops a cosmological singularity and breaks
down while the holographic dual field theory, a sensible Hamiltonian
quantum system typically subjected to a time-dependent gauge coupling,
can potentially be addressed in the vicinity of the singularity. The 
bulk string theory on $AdS_5\times S^5$ (in Poincare slicing) with 
constant dilaton (scalar) is deformed to
$ds^2 = {1\over z^2}({\tilde g}_{\mu\nu}dx^{\mu}dx^{\nu}+dz^2) + d\Omega_5^2,$\
with\ ${\tilde g}_{\mu\nu}, \Phi$ functions of $x^\mu$ alone ($\Phi=\Phi(t)$ 
or $\Phi(x^+)$ then gives time-dependence). This is a solution if\ 
${\tilde R}_{\mu\nu} = {1\over 2}\del_{\mu}\Phi\del_{\nu}\Phi ,\ \ 
{\tilde\Box}\Phi=0$,\ (where ${\tilde\Box}\equiv 
{} {1\over \sqrt{-{\tilde g}}} \del_{\mu} (\sqrt{-{\tilde g}} 
{\tilde g}^{\mu\nu} \del_{\nu})$):\ these include \eg\ AdS-Kasner, -FRW, 
-BKL (based on the Bianchi classification), etc. In many cases, 
it is possible to find new coordinates such that the boundary metric\ 
$ds_4^2=\lim_{z\ra 0} z^2 ds_5^2$\ is flat at least as an expansion 
about $z=0$. This suggests that the dual is the \Nf\ super Yang-Mills 
theory with the gauge coupling $g_{YM}^2=e^\Phi$ deformed to have 
external time-dependence. It is useful to focus on sources approaching\ 
$e^{\Phi}\ra 0$ at some finite point in time: for instance, a 
coupling of the form\ $g_{YM}^2 \ra\ t^p ,\ p>0$,\ gives rise to\ 
$\ R_{tt}\sim {1\over 2} {\dot\Phi}^2\ \sim {1\over t^2}$ , \ie\ a 
bulk singularity with curvatures and tidal forces diverging near $t=0$.

Analyzing the gauge theory is possible in some cases. While at first 
sight one might imagine the dual in such cases to be weakly coupled, 
this is not the case and interactions are important in general 
\cite{Awad:2008jf}. For 
instance, the gauge kinetic terms ${1\over g_{YM}^2(t)} Tr F^2$ can be
transformed to canonical ones for redefined gauge fields (as in 
standard perturbation theory), but this gives rise to new (tachyonic, 
divergent) mass-terms stemming from the time derivatives of the 
coupling, which ensure that the field variables get driven to large 
values as $t\ra 0$. With gauge kinetic terms ${1\over g_{YM}^2(t)} Tr F^2$, 
it turns out that the gauge theory Schrodinger wavefunctional 
near singularity ($t\ra 0$) has a ``wildly oscillating'' phase for 
$p>1$. Furthermore, the energy expectation value generically diverges 
as\ $\lA H\rA \sim {1\over g_{YM}^2(t)} \lA V\rA$. This suggests that
if the coupling vanishes strictly, $g_{YM}^2=e^\Phi\ra 0$, the gauge 
theory response is singular. Deforming the gauge coupling so that\ 
$g_{YM}^2=e^\Phi$ is small but nonzero near $t=0$ leads to finite but 
large phase oscillation and energy production:\ 
${\dot\Phi}\sim {\dot g_{YM}\over g_{YM}}$\ is now finite so the bulk 
is also nonsingular (but stringy). The eventual gauge theory endpoint 
depends on the details of energy production, but one might expect 
thermalization on long timescales if the sources turn off.

With the gauge coupling $g_{YM}^2=e^{\Phi(x^+)}$ being a function of
lightcone time $x^+$, the physics is quite different
\cite{Das:2006dz,Das:2006pw}: in this case, 
${\tilde g}_{\mu\nu}=e^{f(x^+)}\eta_{\mu\nu}$ in the bulk, which is
engineered to acquire a null singularity at say $x^+=0$ (with
$e^\Phi\sim (x^+)^p$). Redefining $A_\mu$ to absorb the coupling gives
canonical kinetic terms and here the potentially problematic
mass-terms in fact vanish due to the lightlike coupling. With these
new gauge theory variables, the interaction terms become unimportant
as $e^{\Phi}\ra 0$. The near singularity lightcone Schrodinger
wavefunctional then appears regular, suggesting weakly coupled
Yang-Mills theory.  These variables appear to be dual to stringy
objects in the bulk (see also related work on worldsheet string
descriptions of certain null Kasner-like singularities
\cite{knNullws}).  We note potential questions about renormalization
effects however, by \eg\ introducing a ``short-time'' cutoff near
singularity, and studying contributions to the gauge theory effective
action from sufficiently high frequency modes (relative to
$\dot\Phi$).

An interesting point in this discussion has to do with the initial
conditions for the time evolution and eventual cosmological
singularity. While in many cases, the sources can be seen to turn off
in the far past thus suggesting the initial state is the vacuum (for
the \Nf\ SYM theory), this is subtle. The fact that there is a bulk
curvature singularity in the deep interior ($z\ra\infty$)
\cite{Awad:2007fj} and the observation that the deformations are
constrained (\eg\ ${\tilde R}_{\mu\nu} = {1\over 2}\del_{\mu}\Phi
\del_{\nu}\Phi ,\ {\tilde\Box}\Phi=0$) suggest that in fact the
initial conditions are constrained or fine-tuned. Exploring this
further turns out to be more fruitful in a related context, that of
similar deformations in de Sitter space.

\section{de Sitter boundary deformations and $dS/CFT$}

\begin{figure}[h]
\includegraphics[width=5.5pc]{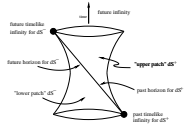}\hspace{2pc}
\begin{minipage}[b]{27pc}\caption{\label{figdS}{\small deSitter space 
$dS_{d+1}$ in the planar coordinates foliation has the metric
$ds^2 = {1\over\tau^2} \left[-d\tau^2 + \delta_{ij}dx^idx^j\right]$: 
this covers half the space ($dS^{\pm}$), and $\tau\ra\pm\infty$ 
with $x^i$ fixed corresponds to past/future timelike infinity, while 
light rays $x^i\sim\tau$ give the horizon.}}
\end{minipage}
\end{figure}
In the following, we will study certain deformations of deSitter
space, containing cosmological (Big-Bang or -Crunch) singularities.
Consider de Sitter space $dS_{d+1}$ in the planar coordinates foliation
(Figure~\ref{figdS}), which is a solution to $R_{MN}=dg_{MN}$ (with 
positive cosmological constant), $\tau$ being conformal time. We 
introduce possible deformations of the $d$-dim Euclidean boundary
\be\label{dSdefmns}
ds^2 = {1\over\tau^2} \left[-d\tau^2 + {\tilde g}_{ij} dx^i dx^j\right]\ ,
\ee
where the spatial metric ${\tilde g}_{ij}$ is a function solely of 
the spatial coordinates $x^i$ (\ie\ not involving time $\tau$). We 
see that these are in a sense analytic continuations of 
the AdS-cosmologies described above, although various physical 
features are quite different qualitatively. These metrics have\
$R_{\tau\tau}=-{d\over\tau^2} ,\ 
R_{ij}={\tilde R}_{ij}+{d\over\tau^2}{\tilde g}_{ij}$,\ and are thus 
solutions to Einstein equations $R_{MN}=dg_{MN}$ (with no additional 
matter) if\ ${\tilde R}_{ij}=0$,\ \ie\ the $d$-dim Euclidean metric 
is Ricci-flat. In general, we consider regular ${\tilde g}_{ij}$. We 
then see that the generic spatial metric ${\tilde g}_{ij}$ (even if 
regular) gives rise to singularities at $|\tau|\ra \infty$\ due to 
diverging\ 
$R^{ABCD}R_{ABCD}\sim \tau^4{\tilde R}^{ijab}{\tilde R}_{ijab}+\ldots$\
($|\tau|\ra\infty$ is the analog of 
$z\ra\infty$ in the $AdS$ interior). We will say more on this below.

Now consider $dS_{d+1}$-deformations sourced by a background scalar 
field $\phi$. If the scalar has purely spatial dependence $\phi(x^i)$,
this is a solution if the $d$-dim part is an Einstein scalar system,
\be\label{ricciscalarEOM}
{\tilde R}_{ij}={1\over 2} \del_i\phi\del_j\phi\ , \qquad
{\tilde \Box} \phi = 0\ ,
\ee
with\ ${\tilde \Box}={1\over\sqrt{\tilde g}}
\del_i({\tilde g}^{ij}\sqrt{\tilde g}\del_j)$.
This scalar is non-dynamical and could represent nontrivial initial or 
final conditions sourcing the deformation of the spacetime.
Using (\ref{ricciscalarEOM}), these solutions have the invariants\ 
$R=d(d+1) + {\tau^2\over 2} ({\tilde\del}\phi)^2$,\ 
$R_{AB}R^{AB}  
=d^2(d+1)+{3\tau^2\over 2}({\tilde\del}\phi)^2+{\tau^4\over 4}
(({\tilde\del}\phi)^2)^2$,\ where 
$({\tilde\del}\phi)^2={\tilde g}^{ij}\del_i\phi\del_j\phi$. 
Thus these invariants diverge at early/late times $|\tau|\ra \infty$, 
sourced by the scalar (even if the scalar energy density itself is 
finite).  

We note now that purely gravitational $dS_4$-deformations do not exist:
${\tilde R}_{ij}=0$ has only trivial solutions, pure 3-dim gravity 
being trivial. Nontrivial $dS_4$-deformations require a nontrivial 
source, \eg\ the scalar field $\phi(x^i)$ above\footnote{For 
$dS_5$-deformations (and higher 
dimensions), there are nontrivial solutions to ${\tilde R}_{ij}=0$, 
\eg\ 4-dim Ricci-flat spaces, which include ALE spaces (noncompact 
$\BC^2/\BZ_N$ singularities and their complete resolutions which are 
smooth) and (compact) K3-surfaces. The $dS_4$-deformation above has 
been found starting with the spatial metric\
$d\sigma^2 = dr^2 + (f(r))^2 d\varphi^2 + h(r) dz^2 ,\ \phi=\phi(r)$.\ 
$C\neq 0$ ensures regularity at $r=0$; also $\phi$ is regular for 
large $r$.}:
\eg\ consider the $dS_4$-deformation\\
\qquad\qquad\qquad
$ds^2 = {1\over\tau^2} \left[-d\tau^2 + dr^2 + (r+C)d\varphi^2 + 2(r+C) dz^2
\right]\ , \quad e^\phi={1\over r+C}\ ,$\\
with\ $R=12+{\tau^2\over 2(r+C)^2} ,\ 
R_{\mu\nu}R^{\mu\nu}=36+{3\tau^2\over (r+C)^2}+{\tau^4\over (r+C)^4}$,\ 
and $R_{\mu\nu\rho\sigma}R^{\mu\nu\rho\sigma}$ has similar structure.
Then for finite $r$, these invariants diverge as $|\tau|\ra \infty$, 
signalling a singularity at past/future timelike infinity.
Light ray trajectories include geodesics of the form\ $\tau=r$, 
reaching the past horizon as $|\tau|\ra \infty$, and here these 
invariants are finite. If higher order invariants are also finite, 
this would give a spacelike singularity at $|\tau|\ra \infty$.

It is interesting to compare these solutions with conventional
investigations of cosmological perturbations about de Sitter space,
\eg\ \cite{Maldacena:2002vr}. In an initial value formulation, 
the metric family\ $ds^2=-N^2dt^2+h_{ij}dx^idx^j$, with $N(t)$ the 
lapse function and $h_{ij}$ the spatial metric, has the action\
$S = \int d^4x\sqrt{h} N\ [ K_{ij}K^{ij} - K^2 + {\tilde R}^{(3)} 
- 2\Lambda + {1\over 2}N^{-2} {\dot\phi}^2 
- {1\over 2}h^{ij}\del_i\phi\del_j\phi] ,$\
with $R^{(3)}$ the 3-curvature, the extrinsic curvature being\ 
$K_{ij}=-{1\over 2}(\nabla_in_j+\nabla_jn_i)=-{1\over 2N}\del_\tau h_{ij}$\ 
($K=h^{ij}K_{ij}$, and $n$ is a unit normal to $t=const$ surfaces).
The action for small gravitational fluctuations\ 
$h_{ij}= a^2(t)(\delta_{ij}+\gamma_{ij})$ becomes\ 
$S\sim \int d^4x (N a^d {1\over 4N^2} (\dot\gamma_{ij})^2 - N a^d {1\over a^2} 
(\del_k\gamma_{ij})^2)$:\ for $dS_4$, using $N=a={1\over\tau}$ 
leads to the familiar Bunch-Davies vacua (after imposing appropriate 
regularity conditions). On the other hand, the 
solutions (\ref{dSdefmns}) arise from the above action in the limit 
where the $\gamma_{ij}$ become essentially time-independent.
The Hamiltonian constraint\
$K_{ij}K^{ij} - K^2 = {\tilde R}^{(3)} - 2\Lambda 
- {1\over 2}N^{-2} {\dot\phi}^2 - {1\over 2}h^{ij}\del_i\phi\del_j\phi$\ 
can be seen to be satisfied for (\ref{dSdefmns}) with\ 
${\tilde R}_{ij}={1\over 2}\del_i\phi\del_j\phi,\ \Lambda={d(d-1)\over 2}$.
The on-shell action for these solutions contains no divergence 
arising from the near singularity region $|\tau|\ra\infty$. Similarly 
the boundary term\ $S_B \sim \int d^3x \sqrt{{\tilde g}} K$\ is also 
regular. This apparent nonsingularity of the wavefunction of the 
universe $\Psi=e^{iS}$ could of course easily be invalidated by 
possibly singular higher derivative terms: one might expect stringy 
effects are important.

From the point of view of $dS/CFT$ \cite{Strominger:2001pn}, 
using the discussion in \cite{Maldacena:2002vr}, 
we expect that the wavefunction of the universe (related to the 
partition function for the dual field theory) is effectively
an analytic continuation of that for similar solutions in Euclidean
$AdS$ as are correlation functions (while observables such as n-point 
functions of bulk fluctuations are not). 
The precise analytic continuation from Euclidean $AdS_{d+1}$ to 
$dS_{d+1}$ (in planar coordinates) is obtained by\
$z\ra -i\tau ,\ \ R_{AdS}\ra -iR_{dS}$. For the deformations we are 
considering, in the limit of small deformations, we see that 
sources are turned on for the operators dual to bulk graviton and 
scalar $\phi$ modes. This leads us to guess that the dual CFT lives 
on the space ${\tilde g}_{ij}$ and has been 
deformed by a source for the operator ${\cal O^\phi}$\ (although for 
our solutions, the deformations are not small). In this perspective, 
these solutions are dual to corresponding deformations of the
Euclidean CFT dual to de Sitter space.

Correlation functions for operators dual to bulk scalar modes can be
calculated by differentiating w.r.t. the boundary sources. 
For instance, by using a Fourier decomposition in terms of 
eigen-modefunctions on ${\tilde g}_{ij}$ and evaluating the
action, it can be seen that the momentum space 2-point function in
these $dS_4$-deformed theories is just as in $dS_4$. The holographic 
stress tensor can be calculated using the usual counterterm 
prescriptions for these theories using \cite{balasub0110108,Awad:2007fj}, 
giving (on one of the $dS$-patches, Fig.~\ref{figdS})\\
\qquad\qquad 
$T^{ij} = {1\over 8\pi G_{d+1}} \left[K^{ij} - K h^{ij} + (d-1) h^{ij}
+ {1\over 2} G^{ij} - {1\over 4}\del^i\phi\del^j\phi 
+ {1\over 8} h^{ij} (\del\phi)^2 \right]$.\\
For the present $dS_{d+1}$-deformations, with boundary metric\
$h_{ij}={1\over\tau^2} {\tilde g}_{ij}$, the extrinsic curvature is\ 
$K_{ij}=h_{ij}$, so that the stress tensor vanishes identically, using 
${\tilde R}^{ij}={1\over 2} \del^i\phi\del^j\phi$.

Recall that if a source for a bulk field is turned
on, we generically expect a nonzero 1-point function for the dual
operator as a response to the source. Here with the metric source 
${\tilde g}_{ij}$, we expect\ $\lA T_{ij}\rA\neq 0$: this is at 
variance with the vanishing stress tensor above.
In fact requiring that the holographic stress tensor vanishes 
(relatedly, trace anomaly encoding conformality of the Euclidean CFT 
on the curved space) gives the conditions\ 
${\tilde R}^{ij}={1\over 2} \del^i\phi\del^j\phi$.

Now consider the ($\tau\sim 0$) Fefferman-Graham expansion for
an asymptotically locally de Sitter spacetime\
$ds^2 = -{d\tau^2\over\tau^2} + {1\over\tau^2}
\big[g_{ij}^0(x^i)+\tau^2g_{ij}^2(x^i)+\ldots\big] dx^idx^j$. 
With the leading source $g_{ij}^0$, the subleading coefficients 
$g_{ij}^2,g_{ij}^4,\ldots$ are generically nonzero \cite{skenderis}, 
and encode information about the state of the
dual CFT. Consider then the deformations above as applied to the 
lower patch $dS^-$: we specify initial conditions for the bulk 
metric and equivalently, the initial conditions/state for the dual 
Euclidean CFT. In the present case, we have seen that in fact 
only $g_{ij}^0\equiv {\tilde g}_{ij}\neq 0$ with $g_{ij}^n=0,\ n>0$:
in fact requiring that the higher order coefficients vanish leads to 
our solutions above. Using the small-$\tau$ Fefferman-Graham 
expansion for both metric and scalar\
$\phi=\tau^{(d-\Delta)/2} (\phi^0+\tau^2\phi^2+\ldots)$, 
we need to solve\ $R_{MN}=dg_{MN}+{1\over 2}\del_M\phi\del_N\phi$\ 
iteratively: this gives\\
\qquad $g_{ij}^2\sim R_{ij}^0-{1\over 2}\del_i\phi\del_j\phi
-{1\over 2(d-1)}\big(R-{1\over 2} (\del\phi)^2\big) g_{ij}^0$ :\qquad 
$g^2_{ij}=0\ \ \Rightarrow\ \ R_{ij}^0={1\over 2}\del_i\phi\del_j\phi$,\\
 (for a massless scalar $\Delta=d$) also implying the higher order 
coefficients vanish. Likewise, with 
$\Box^0$ being the Laplacian w.r.t. $g^0_{ij}$, we also obtain\ 
$\phi^{(2)}\sim \Box^0\phi^0$: thus $\phi^{(2)}=0$ implies $\Box^0\phi^0=0$.
The vanishing of the stress tensor is also related to these subleading 
coefficients.

Our solutions have constrained these subleading pieces of the metric
and scalar in the small $\tau$ expansion to vanish. These conditions
on the $g_{ij}^n,\ \phi^n,\ n>0,$ are highly non-generic and appear to
be nontrivial constraints on the CFT state.

{\bf Discussion:}\ We have argued that these deformations of $AdS$ and
$dS$ are constrained from a Fefferman-Graham perspective, leading to
certain singular structures. This suggests turning on the subleading
coefficients in appropriate fashion towards de-singularizing them: we
hope to explore this further. We believe these arguments also apply 
to the solutions in \cite{Balasubramanian:2010uk}, which, although 
quite different in interpretation, are related to the $AdS$ null 
solutions (sec.~1 above) by coordinate transformations 
\cite{Awad:2007fj}: here diverging tidal forces (with finite 
curvature invariants) lead to the large-$z$ singularity 
\cite{Horowitz:2011gh}.

\vspace{1mm}

{\footnotesize \noindent {\bf Acknowledgements:} 
I thank the ICGC2011 organizers for a stimulating 
conference. Sec.~2 on $dS$-deformations is based on work in progress 
with Sumit Das. I also thank S. Minwalla, M. Rangamani and S. Trivedi 
for useful conversations.
This work is partially supported by a Ramanujan Fellowship, DST, Govt 
of India.}

\vspace{-0.8mm}

{\tiny

}

\end{document}